\documentclass[aip,apl,preprint,showpacs]{revtex4-1}
\usepackage{graphicx}
\usepackage{graphics}

\begin{document}
	
	\title{Domain wall induced modulation of low field H-T phase diagram in patterned superconductor-ferromagnet stripes}
	
\author{Ekta Bhatia}
\affiliation{School of Physical Sciences, National Institute of Science Education and Research (NISER), HBNI, Bhubaneswar, Odisha, India-752050}
\author{Zoe H. Barber}
\affiliation{Department of Materials Science and Metallurgy, University of Cambridge, 27 Charles Babbage Road, CB3 0FS, UK}

\author{Ilari J. Maasilta }
\affiliation{Nanoscience Center, Department of Physics, University of Jyvaskyla, P. O. Box 35, FI-40014, Jyv\"{a}skyl\"{a}, Finland}

\author{Kartik Senapati}
\email[Kartik Senapati: ]{kartik@niser.ac.in}
\affiliation{School of Physical Sciences, National Institute of Science Education and Research (NISER), HBNI, Bhubaneswar, Odisha, India-752050}

\date{\today}

\begin{abstract}
We present a systematic study of the magnetic domain wall induced modulation of superconducting transition temperature (T$_c$) in Nb/Ni bilayer stripes. By varying the thickness of the Ni layer from 20 nm to 100 nm we have been able to measure the low field T$_c$-H phase diagram spanning the N\'{e}el domain wall and Bloch domain wall range of thicknesses. Micromagnetic simulations confirmed a stronger out-of-plane stray field in the Bloch domain walls compared to the N\'{e}el walls. A suppression in T$_c$ was observed in the magnetization reversal region of the Ni film, the magnitude of which followed linearly to the strength of the out-of-plane stray field due to the domain walls. The upper limit of the magnitude of domain wall stray field was roughly estimated by comparing the T$_c$ of the suppressed region of H-T$_c$ phase diagrams with the unaffected part of the H-T$_c$ curve. With Bloch domain walls a change in T$_c$ of more than 60 mK was observed which is much more compared to the earlier reports. We believe that the narrow stripe geometry of the bilayers and the transverse external field maximized the effect of the domain walls in the Ni layer on the overlying superconducting film, leading to a larger change in T$_c$. This observation may be useful for domain wall controlled switching devices in superconducting spintronics.  
	
\end{abstract}
	
\maketitle

The field of superconducting spintronics has attracted wide research interest in recent years \cite{1,2}. One of the reasons is the additional spin degree of freedom provided by the Cooper pairs to the spintronic devices. Since such devices inevitably contain co-functioning S and F components, new interesting phenomena such as $\pi$-phase superconductivity \cite{3}, spin-triplet supercurrent \cite{4}, odd-frequency pairing \cite{5} and long-range magnetic proximity effects \cite{6} emerge out of the natural competition between superconducting and ferromagnetic orders. These effects have been well studied in a variety of devices over the past years\cite{7,8,9,10}. In the context of superconducting spintronics, the effect of magnetization dynamics of ferromagnetic components on the superconducting components is also a very pertinent question.

Typically, magnetization reversal of one or more ferromagnetic components is the key functional aspect of spintronic devices. During the magnetization reversal process, the domain walls in ferromagnets produce stray fields which can alter the properties of a proximal superconducting layer and, therefore, may affect the overall device properties. Prior experimental observations of the superconducting spin switch effect\cite{11}, domain wall superconductivity\cite{12,13} and triplet superconductivity\cite{4} are qualitatively consistent with theoretical predictions of F/S proximity effects involving non-uniform ferromagnets\cite{14,15}. Stray fields invariably accompany inhomogeneous magnetization distributions such as domain walls and sample edges. They can suppress superconductivity by the classical orbital effect or by dissipative vortex motion. Thus, a definitive comparison between theory and experiment in superconducting spintronics is problematic without considering the exact strength of stray field.  Therefore, it is an important parameter to know in proximity effect based or domain state dominated superconducting spintronic devices \cite{16,17,39,40}. However, no direct quantification of the extent to which a superconducting layer is affected by these stray fields, are available in the literature. Some earlier reports have, however, measured the stray field of magnetic thin films using electro-optic studies\cite{18}, photo-emission electron microscopy\cite{19}, magnetic force microscope (MFM)\cite{20}, quantitative MFM \cite{21}, and magnetic transmission X-ray microscopy\cite{22}. 

In superconducting-spintronics devices, the domain structure and hence the stray field of domain walls may be modified below the superconducting transition temperature. In such embedded magnetic layers, there is no direct way of quantifying the stray magnetic field. However, the effects of such stray fields on various superconducting multilayer structures have been explored in the literature. Steiner et al.\cite{23} studied the role of stray fields in an exchange-biased system of the type Fe/Nb/Co/CoO and in Fe/Nb bilayers. Hu et al.\cite{24} reported the stray field and the superconducting surface spin valve effect in La$_{0.7}$Ca$_{0.3}$MnO$_{3}$/YBa$_{2}$Cu$_{3}$O$_{7-\delta}$ bilayers. Curran et al.\cite{16} have imaged the stray fields at the surface of
Nb/Ni multilayer samples at various temperatures using the High resolution scanning Hall microscopy (SHM). Yang et al.\cite{25} have reported the modulation of superconductivity by the stray field of Bloch walls in Nb/Y$_{3}$Fe$_{5}$O$_{12}$ hybrids. In this context, we have quantified the stray field of N\'{e}el domain walls and Bloch domain walls of nickel films in Nb/Ni bilayer stripes below the superconducting transition temperature. For this purpose, we have carefully measured the resistive transition temperatures of lithographically patterned narrow channels of Nb/Ni bilayers, as a function of an in-plane applied magnetic field. We observed a systematic variation of suppression in the low field T$_c$ of Nb/Ni stripes as a function of the thickness of the underlying Ni layer. The observed suppression of T$_c$ gives a direct measure of the strength of domain wall stray field, using the standard BCS type H-T phase diagram. The strength of the out-of-plane stray field of Bloch domain walls was found to be much larger compared to N\'{e}el domain walls in the buried nickel film, below the superconducting transition.

A series of Nb-Ni bilayer thin films was prepared at room temperature in a vacuum chamber with base pressure in the range of $10^{-9}$ mbar, using dc-magnetron sputtering of high purity($99.999\%$) niobium and nickel targets on cleaned Si-SiO$_2$ substrates. The thickness of the bottom nickel layer was varied from 20 nm to 100 nm with steps of 20 nm, while the thickness of the top niobium layer was kept fixed at 55nm$\pm$5nm which is above the coherence length of niobium ($\sim$40nm)\cite{41} in all cases. Films were then patterned into narrow stripes of width 3 micron using a combination of electron beam lithography, reactive ion etching and chemical etching techniques. Nb layer, outside the track region was etched with a 100 watt CF$_4$ plasma in an Oxford RIE system. The Ni layer was etched chemically with a dilute commercial Nichrome etchant from Aldrich.   Transition temperatures were found by electrical transport measurements performed in a standard four probe geometry. Superconducting transition temperatures were measured in the presence of an in-plane applied magnetic field along the width of the stripe. For each measurement, the films were saturated by applying a field of 4000 Oe and then ramped to the measurement field value at a temperature of 5K. Magnetization measurements were performed in a SQUID magnetometer with magnetic field applied parallel to the plane of the films. 

Fig.1 shows the typical magnetization rotation configuration of domain walls in the Bloch and N\'{e}el wall regimes. Due to the nature of rotation of moment in the domain wall, in the Bloch domain walls one would expect more out-of-plane stray field compared to the N\'{e}el wall. In an S-F bilayer stripe geometry, the superconducting film in the long striped region of the pattern would be maximally affected by the out of plane stray fields of the domain walls in the underlying ferromagnetic film. Depending on the film thickness, any ferromagnetic film may have N\'{e}el domain walls or Bloch domain walls as shown in Fig.1. Typically, the domain wall energy per unit area (the sum of anisotropy, exchange and stray field energy densities) gradually decreases with increasing film thickness for Bloch walls, whereas for N\'{e}el walls the domain wall energy increases with increasing film thickness. Therefore, below a certain threshold value of film thickness (where the N\'{e}el wall and Bloch wall energy densities match), N\'{e}el walls become energetically favorable, whereas at a higher thickness, Bloch walls are preferred energetically\cite{28,29,30}. It has been predicted theoretically that the crossover thickness in nickel films is about 50 nm \cite{31,32}. These domain walls have a different out of plane component of the stray field.

In order to look at the domain structures and to check the out-of-plane stray field component as a function of thickness of the magnetic layer, we have performed 3D micro-magnetic simulations\cite{33} on Nickel films. For these simulations, the x and y dimensions of the samples were kept fixed as 2 $\mu$m and 1 $\mu$m, respectively. The z dimension was varied from 20 nm to 100 nm for different samples. Here, x-axis refers to the direction along the length of stripes, y-axis refers to the direction along the width of the stripes and z-axis refers to the axis transverse to the sample plane. The cell size for simulation was kept as (10, 10, 10) nm in (x, y, z) directions. Magnetic field was directed along the width (y axis) of the stripe in the plane of the film. The values of saturation moment, anisotropy constant, and the exchange constant for the simulations of magnetization in Ni films were taken from the literature\cite{34,35}. Fig.2(a) and 2(b) show the micromagnetic 3D OOMMF simulation images of nickel stripe of 20 nm and 100 nm thickness. The simulations were performed for field values ranging from 300 mT to -300 mT where the saturation field is 100 mT. The simulation images shown in Fig. 2(a) and 2(b) are taken for field values near the coercive field. As we reduce the field from positive saturation, the domain nucleation starts from the edges of the stripe to minimize the demagnetization energy\cite{42}. The domain reversal happens in a similar way in both the images, however, the domain wall region is more pronounced in the panel (b) (Bloch regime) compared to panel (a) (N\'{e}el regime). This implies that domain wall stray field per unit area seen by any overlying film would be larger in case of Bloch walls. OOMMF simulation solves the Landau \& Lifshitz equation for each magnetic field to find the minimum energy configuration of magnetic moments. The demagnetization energy is defined as the integral of $(\mu_{0}/2)(-M.H_{d})dV$ over the sample volume, where $H_{d}$ is the internal 'demagnetizing field' due to the magnetization itself, and depends on the sample shape. For an in-plane applied magnetic field in a thin film, the demagnetization factor is negligible, while the volume integral provides a factor of film thickness. Therefore, we plot the demagnetization energy (E) obtained from the simulations, normalized with the film thickness.  The magnitude of the demagnetizing field $H_{d}$ for a homogeneously magnetized thin film should be zero\cite{36}. Since we have performed the simulations for an in-plane applied magnetic field, the demagnetizing field is expected to be ideally zero. However, we mention here that we have performed 3D OOMMF simulations in which the total thickness of the film is divided into many 2D planes, defined by the cell size (10 nm in this case) along the thickness direction. Therefore, the interaction between different planes brings in the observed demagnetization energy in Fig 2(c), which is a reflection of the out of plane stray field. Although it is not possible to quantify the absolute value of the out-of-plane stray field in this case, a comparison of the strength of the stray fields for different thickness is possible from the plots shown in Fig 2(c). From Fig.2(c), it is clear that the 100 nm Ni film has a much stronger stray field (which maximizes at the coercive field) compared to the thinner films.

After establishing the existence a larger stray field in the Bloch thickness range, in Fig.3 we show the T$_c$-H phase diagram for a patterned Nb/Ni bilayer(55nm/40nm) along with the magnetic hysteresis curve of the same bilayer. While ramping the magnetic field down from the saturation field, domain activity starts at around the field value at which the hysteresis loop opens up, as shown by the dotted lines in Fig.3. A decrease in transition temperature with decreasing magnetic field was observed for the Nb/Ni stripes, in the range of magnetic domain activity in the Ni layer. In fact, the low-field T$_c$ was found to follow the magnetic hysteresis loop, attaining a minimum value at a field roughly matching with the coercive field of the nickel layer. On increasing the magnetic field in the opposite direction, from zero, T$_c$ again recovered to the normal value. During the magnetization reversal process, the out of plane stray field of domain walls locally affects superconductivity along the stripe\cite{12,37,38}, resulting in the observed decrease in superconducting transition temperature. This decrease was maximum at the coercive field, because near the coercive field one would expect the maximum domain wall density, producing a large stray field. In the saturated state, the Ni film behaves as a single domain with minimum domain wall stray field.

  In order to emphasize the change in T$_c$, Fig.4(a) shows the normalized R-T curves at three different fields for the bilayer with 100 nm thick Ni film. T$_c$ has been defined as the temperature at 50$\%$ of the normal state resistance. Clearly, the transition at -300 Oe, which is close to the coercive field of the Ni layer in this bilayer, is lower by $\sim$ 64 mK compared to the transitions at fields of 1572 Oe and -1572 Oe. In Fig.4(b), we show a comparison of the  T$_c$-H phase diagrams of patterned Nb/Ni bilayer stripes with nickel layer thicknesses of 20nm, 40nm, 80nm, and 100nm. We notice that the effect of domain wall stray field, near the low magnetic field region, is minimal in the case of 20 nm thick Ni film, which has N\'{e}el domain walls. This effect indicates a weaker out-of-plane component of the domain wall stray field, as expected for the N\'{e}el walls. We also observe that in the saturation field range the T$_c$-H curves are BCS-like, in all cases. The suppression of T$_c$ in the domain activity regime of T$_c$-H phase diagram can only be due to the stray field generated by the domain walls, in addition to the small external field. On the other hand, the suppression of T$_c$ at higher fields (in the saturation range of Ni films) is due to the external applied field as expected. Since the number of domains and the corresponding domain walls in the FM film follow the magnetic hysteresis loop, the average stray field is a function of applied magnetic field. In the saturation field range, the domain wall stray field becomes negligible and in the coercive field range it becomes maximum. Following  Pati$\tilde{n}$o et al.\cite{39}, the field dependence of H$_s$ can be extracted from the magnetization loop as
\begin{center}
	$H_{s}(H_{ap})=H_{s_{0}}(1-|M(H_{ap})/M_{s}|)$
\end{center}
where $H_{s_{0}}$, M and $M_s$ are the maximum stray field at coercive field, the magnetization, and the saturation magnetization, respectively. In Fig 4(c) we have plotted the calculated stray field using this formalism, in order to emphasize the fact that domain wall stray field is the origin of the suppression of T$_c$ in the low field regime. We have estimated the maximum strength of the stray field by drawing a horizontal line at the minimum T$_c$. The field value where this horizontal line crosses the H-T phase diagram in the higher field range was taken as the estimate of maximum out-of-plane stray field (H$_{s0}$). H$_{s0}$ is an out of plane field whereas the field values plotted in H-T phase diagram are in-plane. Therefore, the H$_{s0}$, estimated by this method, is not an absolute measure of the domain wall stray field, rather it provides an upper limit to the strength of domain wall stray field. The extracted estimate of H$_{s0}$ is plotted with the change in T$_c$ near the coercive field ($\Delta$T$_c$) in Fig 4(d). The fact that the overall $\Delta$T$_c$ scales linearly with H$_{s0}$ indicates that domain wall stray field may be an usable control component of superconducting spintronics. Thus, by tuning the domain walls and switching them either on or off via external magnetic field, superconductivity in the overlying Nb film can be effectively modulated.

In summary, we have studied patterned Nb/Ni bilayer stripes with different thicknesses of nickel layer spanning the range from N\'{e}el domain walls to Bloch domain walls. Low field T$_c$-H phase diagrams of these patterned structures were found to follow the magnetization loop of the underlying Ni layer. In the domain activity region, a reduction in T$_c$ was observed which maximized near the coercive field of the Ni film. This indicated that the observed suppression in T$_c$ is a result of the domain wall induced stray field of the underlying Ni layer. We have estimated and compared the maximum strength of stray field due to the N\'{e}el domain walls and Bloch domain walls using the superconducting transition of the overlying Nb layer. The overall reduction in T$_c$ was found to be much smaller in the case of N\'{e}el domain walls compared to the Bloch domain walls. The relative strength of the out-of-plane stray field due to N\'{e}el domain wall and Bloch domain wall of a plain nickel film was also examined using micro-magnetic simulations. The variation of stray field with thickness was consistent in both the cases. There is no simple way of estimating the local magnetic field of the domain walls in an embedded superconductor-ferromagnet hybrid below the superconducting transition. However, our measurements show that the domain wall stray fields can be used as a control parameter in superconducting spintronics devices. Furthermore, the observed change in T$_c$ of more than 60 mK with Bloch walls is much more than earlier reports. This large magnitude of domain wall induced tuning of T$_c$ may be useful for domain wall controlled switching components in superconducting spintronics. 
	
We acknowledge the funding from National Institute of Science Education and Research(NISER), DST-Nanomission (SR/NM/NS-1183/2013) and DST SERB (EMR/2016/005518) of Govt. of India. KS acknowledges useful discussions on demagnetization energy and domain wall stray field with Dr J. R. Mohanty, IIT Hyderabaad .

\newpage	
	\begin{figure}[b]
	\includegraphics[width=12cm]{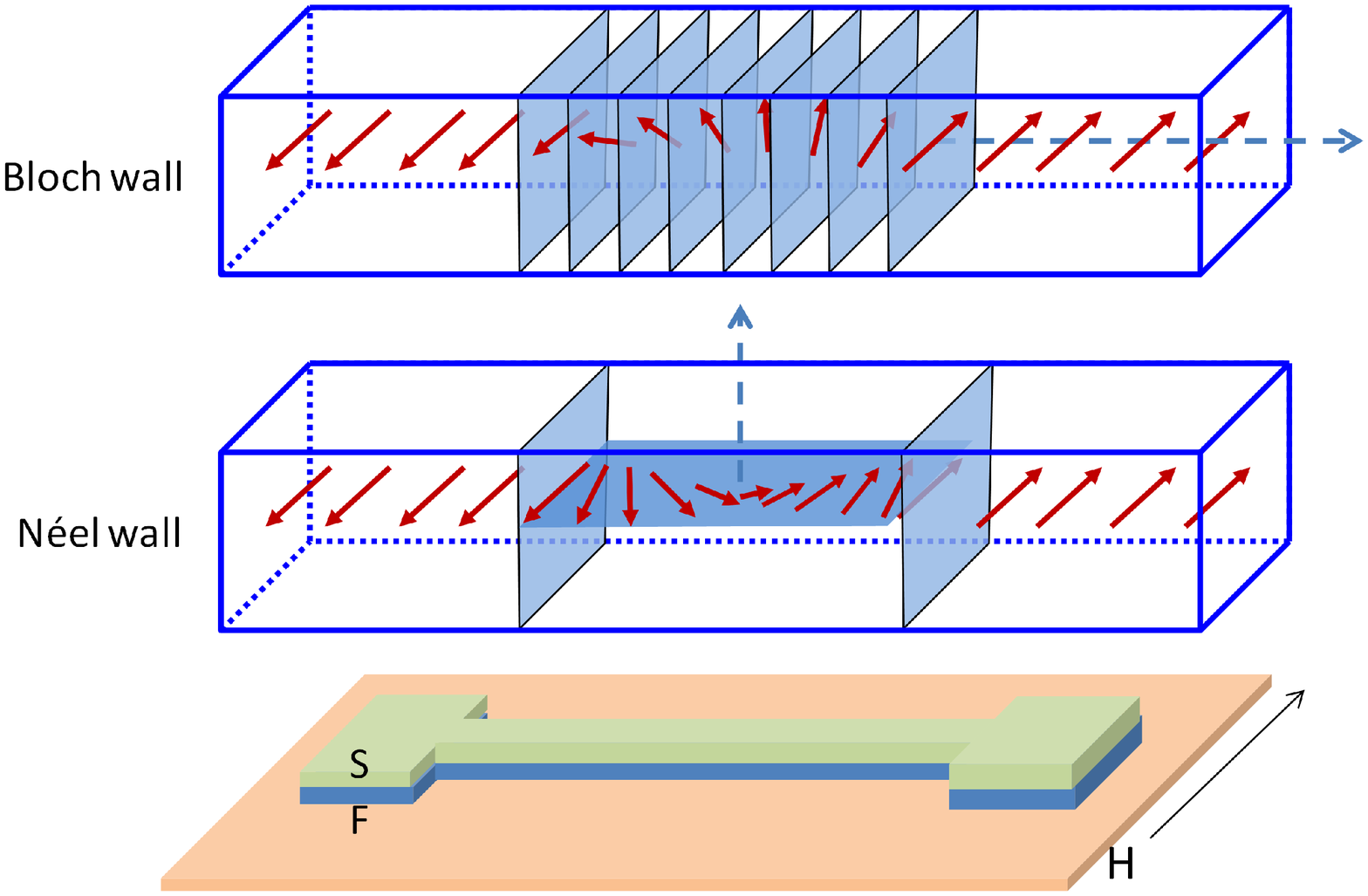}\\
	\caption{Schematic view of magnetization rotation in a N\'{e}el wall and Bloch wall between two domains in a stripe geometry. The dashed arrows show the axis of rotation of magnetization. An overlying superconducting layer, as shown here in the S-F bilayer stripe geometry, would directly sense the out-of-plane stray field of the walls.}
\end{figure}

\newpage

\begin{figure}
\includegraphics[width=12cm]{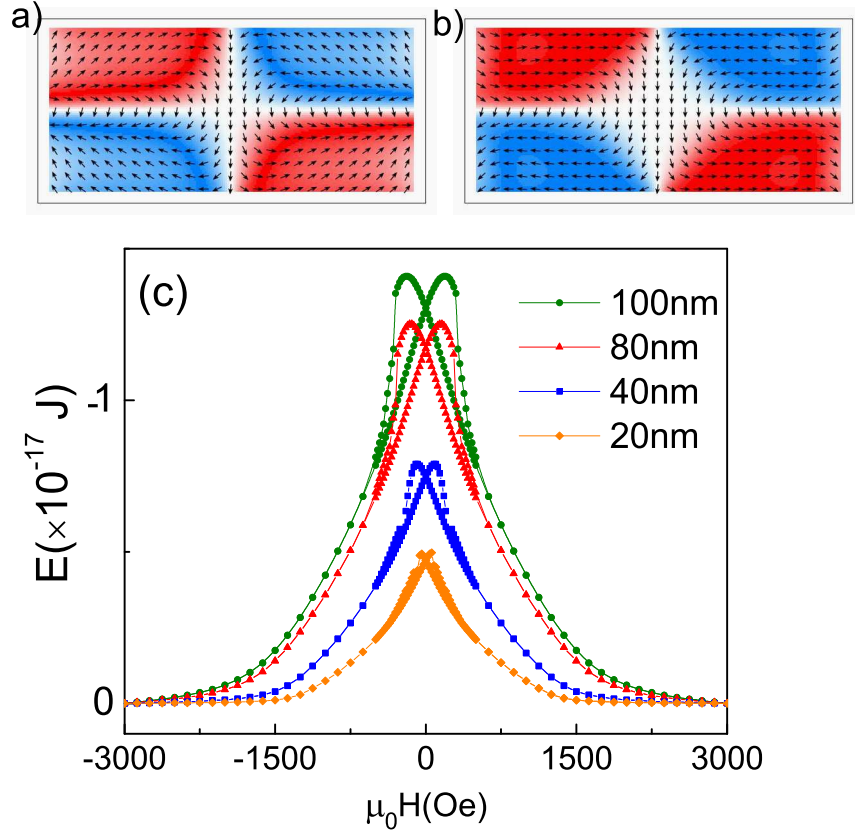}\\
\caption{3D Micromagnetic OOMMF simulation images showing domain structures in Ni thin films of (a) 20 nm (b) 100 nm thickness; these images have been taken for a field value near the coercive field; the red color represents the magnetic moments oriented in one direction while blue color represents the magnetic moments pointing in other direction; the white color presents the domain wall having magnetic moments of changing orientation from red domain to blue domain, (c) Demagnetization energy, extracted from 3D micro-magnetic OOMMF simulation, are compared for films of various thickness.}
\end{figure}

\newpage

\begin{figure}
\includegraphics[width=12cm]{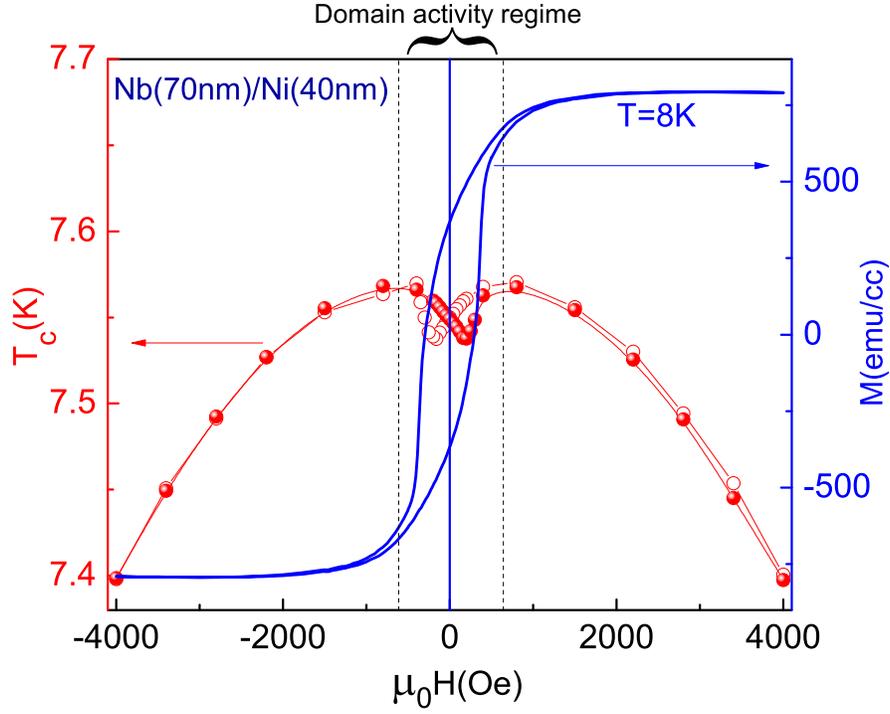}\\
\caption{Transition temperature is shown as a function of in-plane applied magnetic field swept in forward and backward directions, for Nb(55nm)/Ni(40nm) bilayer stripe. Right hand side axis shows the corresponding magnetization loop of the Nb/Ni bilayer at a temperature of 8 K. The minima of the T$_c$ curves clearly match with the coercive field of the Ni layer.}
\end{figure}

\newpage
\begin{figure}
\includegraphics[width=12cm]{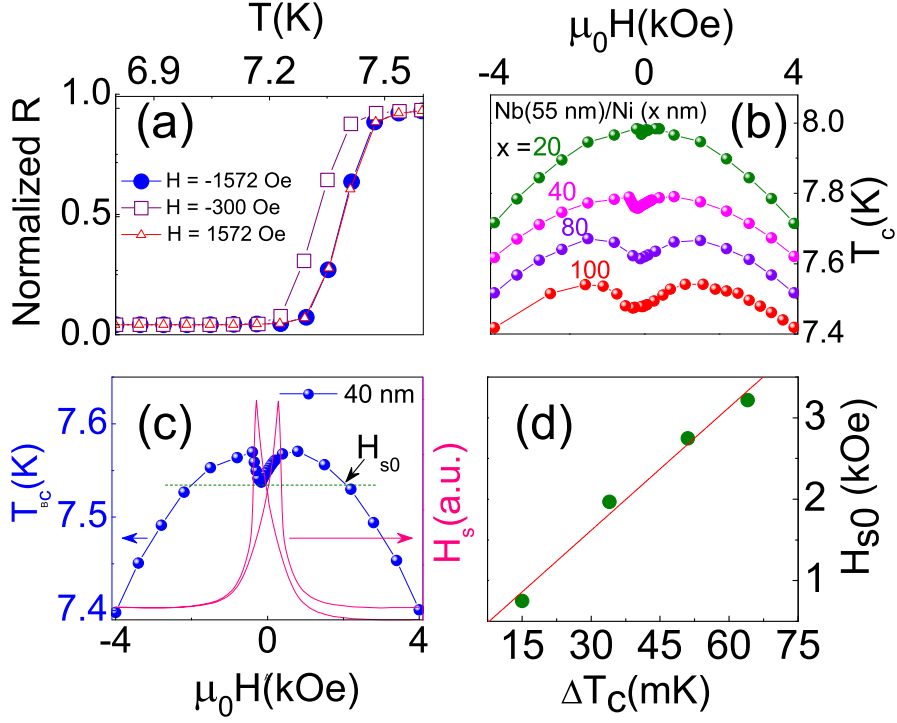}\\
\caption{(a) Normalized resistance vs temperature curves in magnetic fields near the coercive field and saturation field for the Nb/Ni bilayer with 100 nm Ni layer. (b) Comparison of H-T phase diagram of Nb(55)/Ni(x) bilayer stripes with Ni film thickness(x) of 20nm, 40nm, 80nm, and 100nm. (c) H-T diagram of the Nb/Ni bilayer stripes with 40 nm Ni is plotted (on the left hand axis) along with the stray field (H$_s$, on the right hand axis) calculated from the measured  magnetization loop, following Patino et al. \cite{39} as described in the text. The dotted line shows the convention used to find an estimate of the maximum effective out-of-plane stray field (H$_{s0}$) from the H-T diagram of all samples.(d) H$_{s0}$ is compared with the maximum change observed in T$_c$ from T$_c$-H phase diagrams (panel b) of Nb(55)/Ni(x) bilayer stripes.}
\end{figure}

\end{document}